# A Survey on Brain-Computer Interface and Related Applications

[a]Krishna Pai, [b]Rakhee Kallimani, [c]Sridhar Iyer, B. [d]Uma Maheswari
[e]Rajashri Khanai, [f]Dattaprasad Torse

[a,c,e]Department of ECE, KLE Dr. M.S. Sheshgiri College of Engineering and Technology, Udyambag, Belagavi, KA, India- 590008.

[b]Department of EEE, KLE Dr. M.S. Sheshgiri College of Engineering and Technology, Udyambag, Belagavi, KA, India- 590008.

[d]Department of Computer Science and Engineering, Amrita School of Engineering, Bengaluru, Amrita Vishwa Vidyapeetham, KA, India- 560035

[f]Department of CSE, KLE Dr. M.S. Sheshgiri College of Engineering and Technology, Udyambag, Belagavi, KA, India- 590008.

e-mail: krishnapai271999@gmail.com; rakhee.kallimani@klescet.ac.in; sridhariyer1983@klescet.ac.in; b_uma@blr.amrita.edu; rajashrikhanai@klescet.ac.in; datorse@klescet.ac.in

**Abstract**

BCI systems are able to communicate directly between the brain and computer using neural activity measurements without the involvement of muscle movements. For BCI systems to be widely used by people with severe disabilities, long-term studies of their real-world use are needed, along with effective and feasible dissemination models. In addition, the robustness of the BCI systems' performance should be improved so they reach the same level of robustness as natural muscle-based health monitoring. In this chapter, we review the recent BCI related studies, followed by the most relevant applications of BCI systems. We also present the key issues and challenges which exist in regard to the BCI systems and also provide future directions.

## 1. Introduction

BCI uses the brain's power to compute to make use of relatively new technology. Research has been trying to decode the brain's signals since the first discovery of electroencephalography (EEG) a century ago [1]. Until recently, the development of BCIs was thought of as science fiction. BCI system collaborates the brain with an external device that uses signals from the brain for performing external activities such as moving a wheelchair, robotic arm, or a computer cursor. There are four main components of a BCI model, namely, the sensing device, the amplifier, the filter, and the control system. The sensing device comprises a cap consisting of the electrodes which are placed according to the international 10–20 standards [2, 3]. Furthermore, an amplifier could be one of several biological amplifiers available on the market [4], and the research on BCI is geared toward developing a filter and control system that can be applied to brain signals.

When a person thinks of performing any task such as moving a cursor, then in such a case, signals will be generated in the brain which is transferred from the brain to the finger on the computer's mouse via the body's neuromuscular system. As the follow-up step, the finger will move the cursor. In contrast, in BCI, such signals are transferred to an external device where they will be decoded for moving the cursor. As another example, research on BCI also aims to help such people who suffer from issues related to damaged hearing and sight and damaged movement. An estimated 1.5 billion people suffer from neurological and neuroanatomical diseases and injuries worldwide, resulting in movement impairments, which make it difficult to communicate, reach, and grasp independently. A cortical prosthetic system consists of an end effector, which receives a command for a particular action via a BCI that records the cortical activity of individuals who have suffered neurological injuries such as spinal cord injuries, amyotrophic lateral sclerosis, and strokes. In addition, a BCI decodes information pertaining to the intended function. There is a wide range of end effectors in use now, ranging from virtual typing communication systems to robotic arms and hands, as well as functional electrical stimulation for the reanimation of limbs.

It can be invasive in varying degrees, have varying spatial and temporal resolutions, and record a wide range of signals. In BCI applications, such as the low-throughput communication spelling systems [5], EEG, MEG, and fMRI can be used as non-invasive brain imaging technologies. There are some problems associated with these noninvasive BCI approaches, such as the fact that they are often slow (e.g., fMRI), have a low spatial resolution, and are susceptible to being corrupted by external artefacts [6]. Thus, such options are not suitable for complex real-time applications like high-performance communications, tracking multidimensional robotic limbs, or reanimation of paralyzed limbs with coordinated grasps and reaches. A BCI that is invasive, on the other hand, is able to command higher dimensional systems naturally, and restore more complex functions, as a result of its higher resolution and wider transmission bandwidth.

Brain implants are among the most promising and popular technologies for assisting patients with motor paralysis (such as paraplegia or quadriplegia) caused by strokes, spinal cord injuries, cerebral palsy, and amyotrophic lateral sclerosis (ALS) [7]. Similarly, eye tracking can be used to control external devices by paralyzed people, but this tech has numerous drawbacks, as it relies on cameras or electrodes on the face to record eye movements or electrical signals, such as electrooculography (EOG). As a result of BCI, neural commands are delivered to external devices by translating human brain activity into external actions [8-10]. While BCI is most often used to help disabled individuals with motor system disorders, it is also very helpful to those with healthy motor systems as well as the elderly. The development of intelligent, adaptive, and rehabilitative

BCI applications for adults and geriatric patients will enhance their relationships with their families, improve their cognitive and motor skills, and help with household tasks. BCIs are generally regarded as mind-reading technologies, but this does not hold true in most cases. As opposed to mind readers, BCIs provide the user with control by using brain signals rather than muscle movements, so they don't extract information from unknowing or unwilling subjects. A brain-computer interface (BCI) and a user are thus working together through training sessions which involve the user generating brain signals that inform the BCI of the intended action, and the BCI converting the signals into instructions that the output device is supposed to carry out.

As per the aforementioned, the research community faces numerous challenges in the implementation of the BCI devices. Specifically, it is required that the electrodes and the surgical methods used in the BCI process are minimally invasive which has resulted in much research focus on the non-invasive methods of brain-computer interfacing.

In this chapter, we survey the recent research on BCI and its related applications. The related works are detailed in **Section 2**. **Section 3** discusses the most relevant applications of BCI. In **Section 4**, we highlight the main challenges in regard to BCI and propose the relevant directions. Finally, **Section 5** concludes the chapter.

## 2. Related Works

Any human being usually produces a wide range of signals at any point in time which come from the eyes, ears, nose, and other sensory organs present in the body. These signals travel to the brain via the nervous system. The cerebral lobes play an important role for humans or animals when understanding perception, thoughts, language, and memory, and thus EEG sensors, NIRS detectors, etc, are used to acquire neural signals. With the help of these signals, the brain activity of the human or an animal is understood via brain activity measurement algorithms. Fig. 1 demonstrates the detailed BCI interface from which it can be seen that during the process of signal acquisition, pre-processing of the signal is carried out which includes filtering, sampling and artefact removal [11]. Using these pre-processed data, the feature extraction process is carried out following which, classification algorithms or CNN classifiers can be used to understand the neural controls which are required for various applications such as medical gaming education, etc. Further, these neural controls are provided as sensory feedback back to the brain in view of understanding whether the activity which was carried out by the neural control is as desired or not. In this regard, our survey in this chapter is mainly focused on the various types of classification algorithms / CNN classifiers which can/have been implemented for the processing of neural signals, and therefore aid in obtaining the neural controls.

The EEG data which is obtained from the brain in the form of neural signals have multiple channels. The authors in [12] have used single-channel EEG data for developing a prototype using the Internet of things (IoT) and BCI technology. The prototype is developed using the MATLAB software where the classifiers are trained using the Weighted K-Nearest Neighbor Algorithm (Wk-NN). An Arduino microcontroller is used as the hardware platform. The authors have demonstrated that, towards the end of prototyping, a low cost and highly accurate system can be guaranteed which can control the environment. The authors are also able to fetch and provide data to the cloud through IFTTT. The classifiers used in this article are the cognitive state classifier and event-related potential ERP classifier using the Wk-NN algorithm. An accuracy of 95% with 3100 observations per second and an accuracy of 98.3% with 1800 observations per second is achieved by opting for the cognitive state classifier and ERP classifier, respectively.

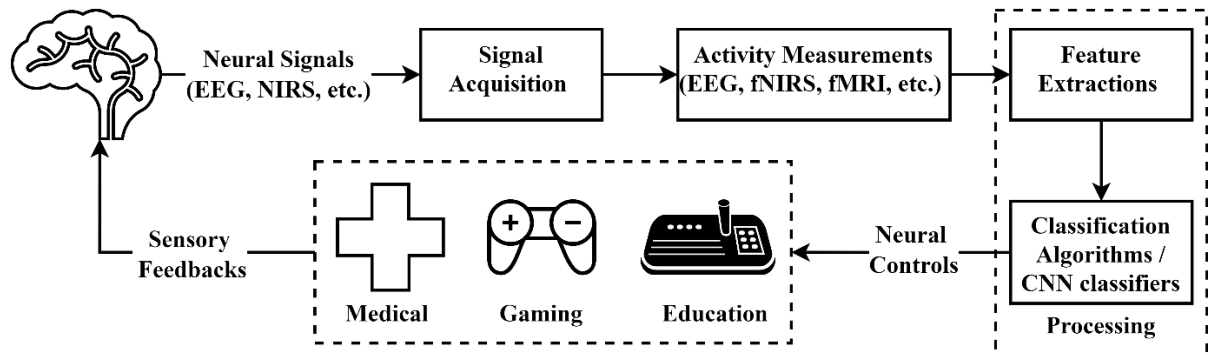

Figure 1: Block diagram of BCI interface.

In general, the classifier algorithms depend on the features which are abstracted from the data that has been collected from the neural signals. There are multiple feature extraction techniques available in the literature, and one such feature extraction technique is the flexible analytic wallet transform (FAWT) method. In [13], the authors have used the FAWT technique to divide the EEG signals into their sub-bands thereby, extracting the movement-based features. Subspace KNN is one of the best classification methods which is known to have reached an accuracy level of 99.33%. The authors have also used other classification methods such as Support vector machines (SVM), decision trees, Linear Discriminant Analysis (LDA) and standard KNN, which have obtained a large spectrum of results. Specifically, an accuracy of 95.72%, 92.8%, 91.79% and 81.1% is obtained by SVM, standard KNN, decision tree and LDA, respectively. The authors have also considered accuracy, specificity, kappa value, f1-score, and sensitivity as the performance parameters.

The authors in [14] have shown that extreme learning machine (ELM) is more efficient than SVM, considering the benchmark performance. They also state that the performance of ELM is highly proportional to the number of hidden nodes, which are also known as the network structure of ELM. Sparse Bayesian ELM based algorithm (SBELM) is shown to exhibit high characterization accuracy on EEG signals. During the comparative study, it is observed that accuracy of 76.3%, 77.1%, 77.8% and 78.5% is obtained via SVM, ELM, BELM and SBELM, respectively. Lastly, the authors have stated that the proposed model can be further improved by adding more distinctive and high-level attributes.

An accuracy level of 95% is obtained using the linear regression method on the equation-based electroencephalogram model by the authors in [15]. The basic machine learning model can be further improved by applying various methods such as systolic matrix multiplication, inversion, and vector multiplication. The proposed algorithm consists of a normal equation that undergoes many mathematical computations to obtain the output. The development environment in the study is built based on docker technology. When deep learning and ELM are combined, in addition to an improvement in a Long Short Term Memory network (LSTMs) and bagging algorithm, a classification model is formed, which is known as LSTMS-B [16]. This new classification model consists of Swish activation. In the study, an intelligent visual classification is observed with an accuracy of 97.13% with 40 classes. Additional image categories are required for a sophisticated technique to distinguish the EEG signals.

In [17], the authors have used the temporal-spatial CNN model which consists of two separate sets of layers namely, the classification layer and feature extraction layer. It is observed that a separate set of data is required for training the feature extraction layers and the classification layers separately. The model provides an overall classification accuracy of 65.7% when spatial filtering is used for feature extraction. The performance which is observed in this study is found to be comparatively better compared to existing classic models. Deep learning models for the BCI interface demands a huge amount of data for high accurate classification results. However, in a general scenario, there is huge data scarcity and hence, the developed model cannot operate at its full potential. To solve this issue, the authors in [18] have proposed a method to generate artificial brain signals which can act as supplementary data along with the actual data. The authors also propose the Deep Convolutional Generative Adversarial Networks (cDCGAN) method for data augmentation which is evaluated on the Convolutional neural network (CNN) model for corresponding classification performance. The CNN classification model is shown to acquire accuracy of 82.86%, 82.86% and 82.14% by employing the actual EEG data, artificially generated EEG data and mixed EEG data, respectively.

The authors [19] have proved the superiority of deep learning-based classification techniques over the existing traditional classification techniques. Five class steady-state visual evoked potential datasets (SSVEP) are used in this study, and a detailed comparison is provided between the CNN and the recurrent neural networks (RNN) using the LSTM architecture. The authors have proved that CNN demonstrates the highest accuracy of classification corresponding to 69.3%, whereas the traditional classification algorithm i.e., SVM with Gaussian kernel, achieves a classification accuracy of 66.9%.

The article [20], provides a comprehensive review of the role of machine learning in BCI, the authors have provided a detailed ML method focusing on mental state detection, state categorization and emotion classification. EEG signal classification, event-related potential signal classification, motor imagery categorization and limb movement classification. The methods such as Common Spatial Pattern (CSP), Principal Component Analysis (PCA), Independent Component Analysis (ICA), Autoregressive (AR) Method, Wavelet Packet Decomposition (WPD) for feature extraction and selection is explored. Detailed Classification methods such as K-Nearest Neighbor (KNN), Linear Discriminant Analysis (LDA), Naive Bayes, Extreme Learning Machine (ELM), Support Vector Machine (SVM), Neural Networks (NNs) are reviewed. The Neural Networks are studied in detail for Multilayer Perceptron (MLP), Artificial Neural Networks (ANN), Convolutional Neural Networks (CNN).

A group of enthusiastic researchers [21], the article is dedicated to the comprehensive review of Development in Emotion detection and classification. The researchers have provided a deep insight in reviewing the published articles focusing significantly on recognition of emotional state based on a specific study performed on the participants/patients at different emotional states using EEG BCI. Researchers all provide a section on trends and challenges in the mentioned field and emphasize the growth of the technology in the next decade.

The authors in [22] provide a complete review of the clinical application of BCI targeting paralyzed patients. The work focused on locked-in syndrome as well as on Completely locked-in syndrome. The work also demonstrates the application of the Brain-machine interface on chronic stroke patients. The results have shown a positive response with the combination of BCI to psychiatric and clinical psychological issues. And the authors mention works further in the improvement in complex behavioural disorders.

The authors in [23] mention different types of neuroimaging BCI methods and provide the merits and demerits of each method. The authors in the article mention the trend of development of the brain-to-brain interface, where individuals are linked with the computer as a mediator. EEG being the popular neuroimaging method, the author describes the other neuroimaging methods such as MEG, which is associated with neuro activity based on magnetic fields. fMRI detects the changes in local cerebral blood oxygen level-dependent signal contrast. NIRS is a method employing a near-infrared spectrum and can penetrate the skull and investigate cerebral metabolism. It is the recent development technique in assessing cortical regional activities. The most recent development is the ultrasound Doppler imaging technique fTCD. The limitation of this method is in terms of penetration.

The authors in [24] worked on EEG data parsed with Discrete Wavelet Transform (DWT) and each multilayer perceptron neural network characteristic are statistically analyzed. The proposed approach presents 98.33 % of accuracy in comparison to other models. The model is proposed to aid in the detection, diagnosis, and classification of epileptic seizures.

The aim of the researchers in [25] is to design a BCI system to extract features and classify the EEG signals accurately by employing Deep Learning Methods. The work is demonstrated on Convolutional Neural networks and Long-term Short-term memory networks. The study is focused on the data obtained from 2 women and 3 men with subjects as healthy, mental and right-handed students.

The authors in the article [26] presented a detailed survey on automated seizure detection. thereby laying a foundation to solve the manual detection method to diagnose epilepsy via manual operations. This study was carried out around various classifiers such as KNN, LS-SVM, Multilayer Perceptron Neural Network (MLPNN), Naïve Bayes (NB) and Random Forest (RF). The highest classification accuracy was achieved to be 97.3% due to an automated system consisting of varied values of Q that decomposes EEG signals. which in turn computes Korsakov and Shannon entropies. Hence detecting the seizures using a classifier named random forest. The other classification accuracies achieved were 86.1%, 95.6%, 92.5% and 84.4% for KNN, LS-SVM, Multilayer Perceptron Neural Network (MLPNN) and Naïve Bayes (NB) respectively.

A highly accurate model [27] for categorisation of normal or seizure conditions for chronic brain disorder is explored. The models are based on tuneable-Q wavelet transform (TQWT) and ensemble empirical mode decomposition (EEMD) algorithms. The entropy parameters and TQWT parameters Have produced an optimum value for high classification performance. The joint method consisting of EEMD-TQWT + RF algorithms have produced and highest classification accuracy of 96.2%.

In Table 1, we summarize the classifier algorithms implemented by various studies including the accuracy which has been achieved.

Table 1. Details of the algorithms used by various studies and the achieved accuracy.

| References | Classifier Algorithm Used | Accuracy |
|---|---|---|
| [12] | Event-Related Potential (ERP) using Weighted k-Nearest Neighbor (Wk-NN) algorithm | 98.3 % |
| [12] | Cognitive State Classifier using Weighted k-Nearest Neighbor (Wk-NN) algorithm | 95 % |
| [13] | Subspace KNN | 99.33% |
| [13] | Support vector machines (SVM) | 95.72% |

| [13] | Standard KNN | 92.8% |
|---|---|---|
| [13] | Decision Tree | 91.79% |
| [13] | Linear Discriminant Analysis (LDA) | 81.1% |
| [14] | Support vector machines (SVM) | 76.3% |
| [14] | Extreme learning machine (ELM) | 77.1% |
| [14] | Bayesian ELM (BELM) | 77.8% |
| [14] | Sparse Bayesian ELM (SBELM) | 78.5% |
| [15] | Linear Regression | 95% |
| [16] | Deep learning + ELM + LSTM + bagging algorithm (LSTMS-B) | 97.13% |
| [17] | Temporal-spatial CNN | 65.7% |
| [18] | Convolutional neural network (CNN) | 82.86% |
| [18] | CNN + Deep Convolutional Generative Adversarial Networks (cDCGAN) | 82.86% |
| [18] | CNN + Mixed Data | 82.14% |
| [19] | CNN/recurrent neural networks (RNN) using LSTM architecture | 69.3% |
| [19] | SVM with Gaussian kernel | 66.9%. |
| [26] | KNN | 86.1% |
| [26] | LS-SVM | 95.6% |
| [26] | Multilayer Perceptron Neural Network (MLPNN) | 92.5% |
| [26] | Naïve Bayes (NB) | 84.4% |
| [26] | Random Forest (RF) | 97.3% |
| [27] | tunable-Q wavelet transform (TQWT) + ensemble empirical mode decomposition (EEMD) + RF | 96.2% |

## 3. Applications of BCI

In this section we details the state-of-the-art applications of BCI. Several studies are underway to enhance current BCI systems by combining multimodal signal acquisition methods. Several studies have demonstrated that fMRI with simultaneous EEG can yield complementary features through the use of the EEG's effective spatial resolution and the EEG's effective temporal resolution [28]. MEG can also work in conjunction with EEG, since it provides information about radially/tangentially polarized sources in cortical subcortical networks, and is able to complement the EEG by adding complementary information [29]. A few studies contend that EEG and MEG can detect subcortical activities; however, skepticism remains regarding their capability to detect brain activities that originate from subcortical areas [30, 31]. It has become increasingly common to combine different signal acquisition methods for improving BCI efficiency in recent years.

In order to translate any brain signal into a command that can be used by a computer or other external device, signal processing combined with machine learning techniques plays a crucial role. Time domain representations of the brain signals include the Fourier transform (FT) and autoregressive models, whereas time-frequency representations include short time FTs and wavelet transforms [32]. When spatial filtering is considered, many inverse models enable the differentiation and projection of actual sources on three-dimensional cortical-subcortical networks [33]. A variety of linear and nonlinear classification algorithms, such as kernel-based support vector machines and linear discriminant analysis can be used to translate the extracted features [34]. A deeper understanding of BCI based on deep learning paradigms is receiving increasing attention from researchers thanks to the remarkable advancements in computational infrastructure over the past decade [35]. This is due to the fact that such systems can evaluate large datasets. The inverse problem must be solved in order to model the cortical sources [33]. Additionally, new methods for localizing sources have emerged in recent years, including wavelet-based maximum entropy over the average that represents EEG/MEG signals as time-frequency contents and then transforms them into spatial representations. Sensors with customized designs are developed for the acquisition of brain signals. There are many forms of neurosensors, including electrochemical, optical, chemical, and biological [37]. Nanowire Field Effect Transistor and other p/n junction devices have demonstrated the feasibility of neuro-sensing techniques in intracellular recordings even in deep brain regions, thanks to major advancements in nanotechnology [38].

BCI systems have also shown the ability to read thoughts in regard to the multiple movements [39-41]. Researchers have used BCI to test the thinking of pigs while running on the treadmill. They were also able to predict what the pig's brain would do next in addition to interpreting the pig's thoughts. Moreover, researchers are carrying out research on the use of brain-computer interfaces to improve communication among people who are paralyzed physically and unable to speak or move. In these situations, BCI enables the person to communicate verbally and in writing. BCI system for brain-to-text transcription also aid the paralyzed people to imagine writing the letters, and these letters appear on the screen. In the experiments, the paralyzed patients are asked to think at a rate of 90 characters per minute, which are then decoded and presented on the screen in less than one second. This demonstrates results which are 80% faster than the typical typing speed on a smartphone screen for a person of the same age range. Even after bein affected for years after paralysis, the motor cortex is still powerful enough to read by a BCI well enough for the typing speed and precision.

In [42, 43], the authors have developed BCI to generate synthesized speech. To obtain the results, electrodes were placed on the surface of the brain to measure the signal and calculate movements of vocal tract. The captured vocal tract movements are then converted to sounds, and the original voice is then converted into synthesized speech. BCIs have also been demonstrated to be used for assiting and restoring the vision of the blind people [44, 45]. A camera, positioned on glass, captures a video image, processes it, and then activates a chip installed in a retina to stimulate the eye. The ultimate goal is to be able to use the implant to transform camera inputs into brain activity. Thus, it presents a novel method of treating blindness which targets the brain directly rather than the eyes. Patients with hearing loss have several options for addressing their hearing loss, including BCIs [46, 47]. It works similar to the microphone which picks up sounds and conversations from various people. Implants are positioned behind the ear or beneath the skin which then transfers the impulses to electrodes which are implanted in the cochlea via a sound processor.

BCIs have also been found to be helpful in assisting the elderly people [48]. Changes occur in the brain and the body as age increases, and elderly people require assistance. Nevertheless, they are generally not provided the assistance they require owing to the expense of the care and treatment which is required. In this case, BCI technology can help the elederly in multiple ways such as, improving their communication, controlling domestic appliances, and strengthening their cognitive abilities. Currently, demantia is very common among adults [49, 50]. BCI s are used to detect early Alzheimer, and are also used to classify and to know the type of Demantia. It is very important to detect early Alzheimer's type of dementia (AD) as it can be treated using medicine itself. Motion sickness and drowsiness decreases the performance of drivers, and on occasions also result in an accident [51-55]. Motion sickness can be predicted by measuring EEG signal received from different areas of brain region. EEG-based BCI systems have shown immense potential in a variety of applications including, post-stroke treatment [56, 57], illness diagnosis [58], emotion identification [59], and gaming [60], have showed great promise for EEG-based BCI systems.

BCI systems also aid in predicting brain tumour [61], epilepsy (seizure disorder) [62], and encephalitis (brain swelling). Further, it is poosible to ensure early detection of Dyslexia using BCI which in turn helps in treating the children and improve their self-confidence [63]. The combination of rtfMRI-EEG BCI systems are used for finding the depression [64], and emotion recognition [65] of patients can also be identified using fMRI BCI system. The BrainArena [66] connects two brains to a football video game using BCIs, and can score by thinking left and right movements. Playing Brainball game twice a week has been demonstrated to improve the students' learning skills. Lastly, people can measure their stress level by playing the Brainball game in which the player with less stress only can move the balls [67].

**4. Issues and Challenges, and future directions**

The BCI technology has gained tremendous attention from the research community including the scientists within the medical and the non-medical fields. The major issues faced within the research domain can be categorized into three sectors namely, neuro-psycho-physiological, technical, and ethical [68].

1. **Neuro-psycho-physiological issues**: Performance of the brain is affected by the anatomy problems including the complexity owing to the genetic issues and structure, diversity of the brain, and the psychological problems such as, anxiety, fatigue and emotional state, stress, and memory which vary from one person to another. These issues have been demonstrated to predominantly affect the performance of BCI.

2. **Technical issues**: The major challenge faced by the BCI system is to select the appropriate components or technology related to the application as it is related to the signal. Selection of the method to acquire the brain signal and then to process it presents a mojor challenge. Another challenge is educating the operator of the BCI system.

3. **Ethical issues**: These are related to the safety of the physical and mental, and emotional state of the user. The user data is highly confidential and needs to be maintained by the system. User consent is another prominent issue related to the BCI system.

In regard to the technical issues, the authors in [69, 70] detail the the significant challenges in developing the BCI. Specifically,

- The Non-Linearity characteristics of the brain signal, with the non-stationarity behaviour of the signal, presents a key challenge. In addition, noise also aids its vital contribution to the challenges of the BCI system.
- Another technical challenge is the brain signal transfer rate. Currently, the BCIs are at an extremely slow transfer rate, and this is a msjor research topic, especially for BCI based on visual stimulus.
- The selection of appropriate decoding techniques, processing and classification algorithms is a challenge to control the BCI system.
- Another critical issue is the lack of balancing between the training required for the accurate function of the system and the technical complexity of decoding the activities of the brain.

- There is a need to focus on the BCI system and provide a systematic approach for a particular performance metric as there are varied performance metrics, and it is a major challenge to select a particular metric for a specific application.

In addition, the areas which need to be further explored include the long-term effects which are not known, technology effects on the life quality of the multiple subjects and their relatives/families, the side effects which are related to health such as, quality of sleep, functioning of the normal brain and memory, and the non convertible alterations which are made to the brain. Further, there also occur multiple legal and social issues which needs to be settled namely, the accountability and responsibililty which is required to be taken in regard to the influence of the BCIs, inaccuracy in the translation of the cognitive intentions, the possible changes in the personality, no being able to distinguish between humans and the machine controlled actions, misuse of techniques during the interrogation by authorities, the capability and privacy of the mind reading, and the mind control and emotion control related issues. In addition, a major are of concern is the legal responsibility which needs to be finalized for scenarios where accidents occur.

Additional challenge which has emerged is in regard to the response of the body to the invasive BCIs which requires the use of implanted micro-electrodes array which come under direct contact with specific neurons within the brain. These electrodes are recognised as foreign bodies which trigger the natural immune system, and these neurons are surrounded by fibrous capsules of the tissue in turn minimzing the signal recording ability of the electrodes, ultimately resulting in the BCIs' unusability. Further, minimizing the power consumption for decreasing the battery size and prolonging the lifespan is another key challenge since there exists a trade-off between the power consumption and the efficient bio-security. In this regard, for ensuring the bio-security, signals are required to be encrypted which increases the power consumption.

Further, the major problem in the implementation of the BCI technology is a lack of efficienct sensor modality which provides safe, accurate, and robust access to the brain signals. Also, the development of such sensors, with additional channels for improving the accuracy and reducing the corresponding power usage is a major challenge. The ethical, legal, and social implications of the BCIs may also slow down, stop, or divert this technology into a completely different path compared to the initial aim. Lastly, over the last decade, technology related to the genetic sequencing technology and modern tools to map have aided the increase in understanding the neuronal firing patterns, and the manner in which they lead to different actions. The brain-interfacing devices are now becoming more sensitive, smaller, smarter, and portable over the time, and future technologies must address the issues which are related to the ease of use, performance robustness, and cost reduction.

## 5. Conclusion

The BCI community is conducting vast amounts of research to provide standardized platforms and to assist the complex and non-linear dynamics that BCI systems encounter. In this chapter, we have reviewed the most recent studies related to the BCI systems. We have also presented and detailed the state-of-the-art BCI systems. Lastly, we have listed the key issues and challenges which exist in regard to the BCI systems followed by some future directions which can enhance the research to be conducted on the BCI systems.